\documentclass[12pt,a4paper]{article}
\usepackage{graphics}
\usepackage{afterpage}
\usepackage{enumerate}

\usepackage[pdftex]{graphicx}

\begin{document}

\title{\bf The study of ground-level ozone in  Kiev \\
and its impact on public health }

\author {A.V. Shavrina$^1$, I.A. Mikulskaya$^2$, S.I. Kiforenko$^2$,\\
V.A. Sheminova$^1$, A.A. Veles$^1$, O.B. Blum$^3$\\[2mm]}
\date{}
\maketitle

\begin{center}{

$^1$ Main Astronomical Observatory,
         National Academy of Sciences of the Ukraine, \\
          27 Akademika Zabolotnoho St.,
          03680 Kiev, Ukraine \\
$^2$  International Research and Training Center of Information
     Technologies and Systems \\
      of the National Academy of Sciences and
      Ministry of Education and Science of Ukraine,\\
      40 Academika Glushkova St., 03680 Kiev, Ukraine \\

$^3$  N.N. Grishko National Botanic Garden of National
       Academy of Sciences of  Ukraine,\\
    1 Timiryazevskaya St., 01014 Kiev, Ukraine \\

}
\end{center}

\begin{abstract}

Ground-level ozone in Kiev for an episode of its high
concentration in August 2000 was simulated with the model of the
urban air pollution UAM-V (Urban Airshed Model). The study of total ozone over Kiev
and its concentration changes with height in the troposphere is
made on the basis of ground-based observations with the infrared
Fourier spectrometer at the Main Astronomical Observatory of
National Academy of Sciences of Ukraine  as a part of
the ESA-NIVR-KNMI no 2907. In 2008 the satellite Aura-OMI data
OMO3PR on the atmosphere ozone  profiles became available.
Beginning in 2005, these 
data include the ozone concentration in the lower layer of the
atmosphere and can be used for the
evaluation of the ground-level ozone concentrations in all
cities of Ukraine. Some statistical investigation of ozone air
pollution in Kiev and medical statistics data on  respiratory
system was carried out with the application of the
``Statistica'' package. The regression analysis, prognostic
regression simulation, and retrospective prognosis of the
epidemiological situation with respect to respiratory system
pathologies in Kiev during 2000--2007 were performed.

\end{abstract}


\section{Introduction}

Simulation of ground-level ozone concentrations in Kiev for an
episode of its high concentration in August 2000 
\cite{Bly02,Sha08_2,Sha10_2} 
is performed using a model of urban air pollution UAM-V 
\cite{Sys95}. 
Aa study of total ozone over Kiev and its concentration changes
with altitude in the troposphere was also carried out using ground-based
observations with the infrared Fourier spectrometer at the Main
Astronomical Observatory of the National Academy of Sciences of
Ukraine (MAO NASU) as a part of the ESA-NIVR-KNMI project
no~2907 ``OMI validation by ground based remote sensing: ozone
columns and atmospheric profiles (2005--2008)'' 
\cite{Sha07,Sha08,Sha10}.  

Ground-level ozone is a highly toxic gas relative to
humans and all living matter. It is formed in photochemical
reactions of precursors exhausted mainly by vehicles
and large industrial plants.  In fact, it is an indicator
of anthropogenic pollution of the studied areas 
\cite{Mik08,Sha08_2,WHO00,WHO05}. 

Ozone was revealed by Schonbein in 1840 and almost immediately
its concentration was began to measure at the Pic du Midi in
France. A long series of mountain observations from 1870 to the
present day showed the increase in background ozone
concentrations in air, from 10 ppb (parts per billion by volume)
to 50 ppb. This last value is now recommended by the World
Health Organization (WHO) as the maximum allowable 8-hour ozone
concentration 
\cite{WHO05}. 
Today, the 2171 ozone monitoring stations  are officially
registered in the monitoring network of Europe 
\cite{air_09}. 
The stations are of
different types: urban, rural and background, including 2111
stations in EU countries. The greatest number of stations are located in
the countries such as France (439), Spain (367), Germany (286),
Italy (235). In the Czech Republic there are 71 stations,
in Poland are 68, in Romania are 27, in Latvia are 6, in Lithuania are 15,
and in Estonia are 7 stations. At the same time in Ukraine, we do
not have any ozone monitoring stations  registered officially in
the aforementioned European network. Our metrological
service does not have calibration instruments for measuring
ozone with gas analyzers. However, ozone pollution in Ukraine
(Kiev, the Carpathians) and its effect on plants was studied in the papers
\cite{Bly06,Bly97} 
since 1996.

Intensive study of the ozone chemistry and harmful effect of
ozone on human health and crops began after 40\% crop losses in
Europe in 1940. Ozone is formed by the reaction:
$\rm O + O_2 + M \equiv O_3 + M$,
where $\rm M$ is molecule-catalyst, such as $\rm N_2$.
In the urban atmosphere the oxygen atom  is formed from nitrogen
oxides, $\rm NO_x (NO_2 + NO)$ by photochemical reaction under
the influence of solar radiation with wavelengths less than
424~nm (a wavelength  less than 290~nm  is delayed in
stratospheric ozone layer at an altitude of  20--35~km above
land): $\rm NO_2 + hv \equiv NO + O$. But again, $\rm NO$ is oxidized by
ozone to $\rm NO_2$, so the accumulation of ozone in this cycle does
not occur. There must be another way of oxidation of $\rm NO$ to
$\rm NO_2$ and it was found. There are  peroxide radicals which are
formed by oxidation of volatile organic compounds both natural
and anthropogenic origin: $\rm NO + RO_2 \cdot \equiv NO_2 + RO \cdot $,
where $\rm R$ is any organic fragments such as $\rm C_2H_5$.

As chemical and toxic substance the ozone has been well studied
\cite{WHO00,WHO05}. 
As for its effect on the  health of the
inhabitants of modern cities, this question
in Ukraine is only beginning to study in contrast to the United States
and European countries (see, e.g., 
\cite{Lip_89,Whi94,WHO00,WHO05,Yang03}.  
In Ukraine this
problem requires the most careful consideration: firstly,
because of the complexities of economic, medical and
environmental conditions prevailing in recent years, and
secondly, because there is a high transformation of cities into
modern metropolises with a huge flows of vehicles. This
phenomenon is new from many points of view and it has not yet
been studied, including health and environmental aspects.

The main goal of this work is to carry out a retrospective study
of the influence of ozone on the human health in  Kiev, the capital 
of Ukraine, since 2000. Unfortunately, we do not yet have the
Health Statistics (HS) data for 2008--2010.

It is known that ozone is air pollutant of the first class
of danger according to the action on human and animal health,
plants and building facilities. The main tagits of ozone action
on human health are the respiratory system (RS) and
cardiovascular system (see, for example, 
\cite{Lip_89,Mik08,Yang03}.  
Our purpose
is the study and validation of forecasting of the
connection between ozone concentrations in Kiev and the state of the
respiratory system of the city population. It is expected  to
use the results of this study to assess the risks of harmful
effects of ozone on the health of the inhabitants of Kiev and
other cities of Ukraine  as at the population and individual
(personal) levels. Note very high summertime concentrations of ozone
in Odessa city (up to 200 ppb), given by the simulation by EURAD
prognostic system of the Rhenish  Institute for Environmental
Research at the University of Cologne (http://www.eurad.uni-koeln.de).

\begin{figure}[t]
\centering
\includegraphics [width=0.95\textwidth, angle=00]{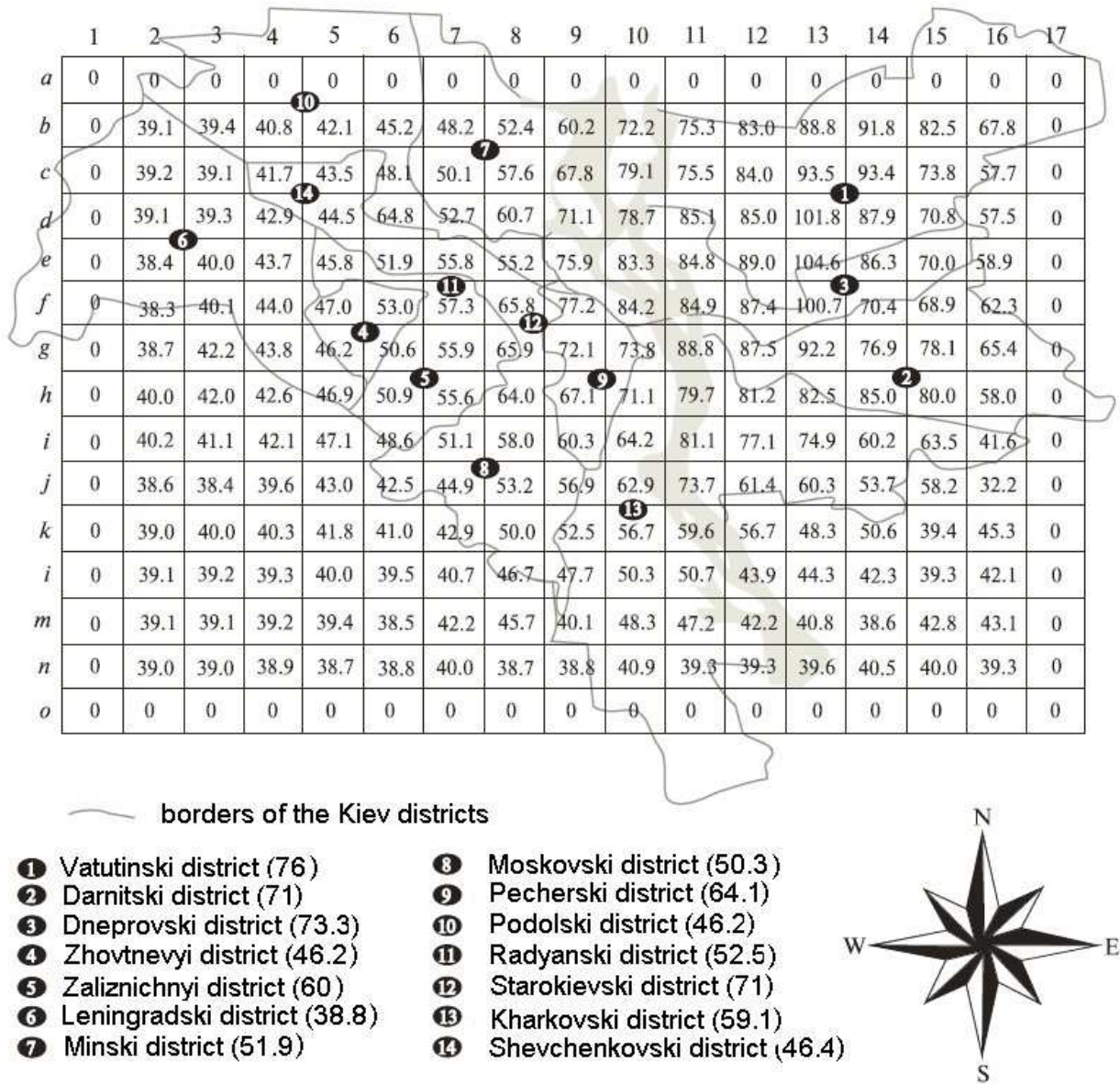}
\caption[]{Distribution of mean maximal ozone concentrations in
Kiev derived from the modelling. Average ozone concentration in
each Kiev district is given in brackets. The ozone
concentrations are given in ppb.}
\end{figure}

\section{Materials and methods}
To perform the planned work we used the statistical data of  the
City Department of Health (CDH) and the Ministry of Health (MH)
for the period 2000--2007 years. The data contains
12 indicators determining the overall and primary
morbidity of ozone dependent RS pathologies for the population
of 14 Kiev districts. These data
related to five socio-age groups of Kiev's inhabitants:
children, teenagers, adults,  working age, and pensioners.
Figure 1 shows the averaged maximum ozone  concentration  for 14
districts of Kiev as well as for Kiev as a whole for the ``ozone
episode'' in August 2000. These data  were obtained  by
the simulation of the ozone pollution taking
into account the ozone formation  processes and  scattering of
ozone forming matter in the surface layer of the atmosphere
over the city 
\cite{Sha08_2,Sha10_2}. 
In the simulation the UAM-V model 
\cite{Sys95} 
was used  taking into account the following
factors: the relief
of the city, weather conditions, the intensity of solar
radiation, the number of volume and point emissions to the
atmosphere by industrial enterprises, the number of vehicles in
the city, the speed of traffic flows.

The satellite Aura-OMI data on the atmosphere profiles
of ozone concentration OMO3PR
(http://disc.sci.gsfc.nasa.gov/Aura/data-holdings/OMI/omo3pr$_-$v003.shtml)
have appeared in 2008. They include ozone data for the lower layer of
the atmosphere beginning in 2005. We performed a comparison of
these ozone profiles for Kiev and the profiles recovered by us from
the infrared Fourier spectrometer observations and modelling
with MODTRAN4 code 
\cite{Sha07,Sha08,Sha10}. 
The first comparison of our retrieved profile
for April 23, 2007 with OMO3PR data showed a significant difference in
tropospheric part of the profiles (Fig.~2a). However, the profile
of new version of OMI data in 2009 is in good agreement with our
profile for the same date (Fig.~2b). This is a good reason for using OMI
data in evaluating ground-level ozone concentrations in all
cities of Ukraine.

Statistical study of available data was carried out  using the
software package ``Statistica'' (e.g.,
http://softnic.ru/soft/programm$_-$4456.html). It included a
correlation analysis of the Kiev district data on ozone
pollution and the RS state data averaged in each district of the
city. In such way we obtained the correlation coefficients between the
compared  values, performed a regression analysis, and
 constructed predictive regression models.

\begin{figure}
\centering
\includegraphics [width=0.8\textwidth, angle=00]{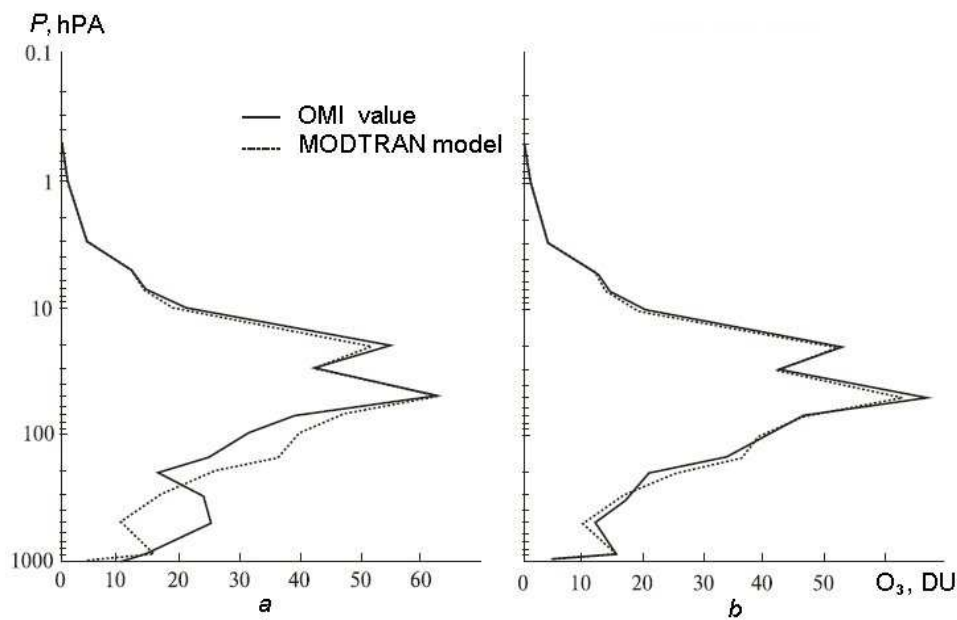}
\caption []{Comparison of the atmospheric ozone profiles derived from  
the OMI data (OMO3PR) of version 2008 (a) 
and version 2009 (b)  with  our atmospheric ozone profile (April 23, 2007) 
retrieved by MODTRAN modelling. Ozone concentrations are expressed 
in the summary quantities for each of the 18 atmospheric layers defined  
by the OMO3PR data.   }
 
\end{figure}

\begin{figure}[t]
\centering
\includegraphics [width=0.85\textwidth, angle=00]{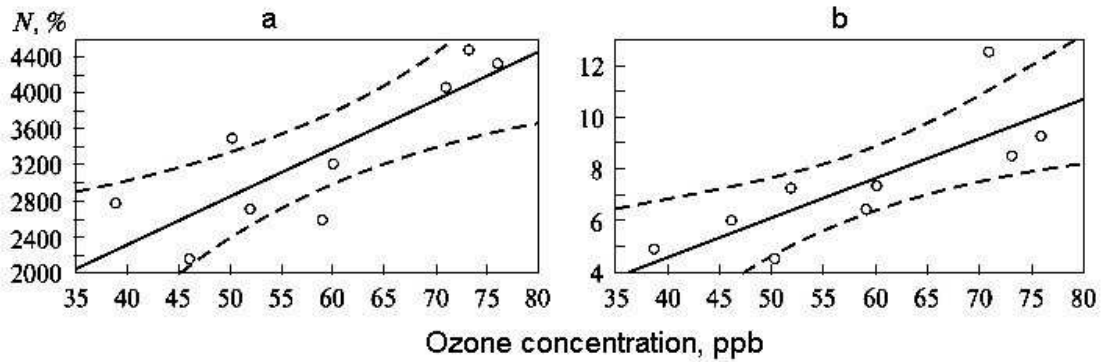}
\caption[]{The prevalence ($N$) of respiratory diseases of the
adult population per 100 000 (a) and asthmatic bronchitis of
children per 1000 children (b) with average ozone concentrations
for the "sleeping" districts of Kiev.   The dashed curves show
the confidence level of 95 percent.}
\end{figure}

\section{An epidemiological study of the ground-level ozone\\ impact
on the RS state of Kiev population in 2000}

The first epidemiological study of the ozone problem in Ukraine
was carried out using our modelling of ozone concentrations in
2000 
\cite{Mik08}. 
At that time,  the administrative structure of Kiev included 14
districts. Each district was served by a network of district
clinics supplying statistics data in Kiev centre ``Medinstat''.
This gave the opportunity for a full statistical comparison of
averaged district RS indicators of the city population and
averaged maximum district ozone concentrations calculated by the
modelling code UAM-V 
\cite{Sys95} 
(Fig.~3). The correlation study showed
for the socio-aged group ``adults'' (it is more than 70\% of the
population) statistically significant relationships emerged in
terms of ``respiratory diseases'' and ``pneumonia''. For
socio-aged group ``children'' a statistically significant
correlation was revealed in terms of ``asthmatic bronchitis''
($r = 0.66$). The correlation between asthma in children group
and peak ozone concentrations was studied, for example, in the
paper  \cite{Whi94}. 

When selecting from the investigated city districts  the nine
suburban ``bedroom'' districts (Vatutinsky, Darnytskyi, Dnieper,
Zhovtnevyi, Zaliznichnyi, Leningrad, Moscow, Kharkov and
Radyanskyi) with their socio-economic, medical, and ecological
features, and with their almost the same population density (five
thousand people per 1~km$^2$), we found a high correlation
between ozone concentrations and medical data. For adults the
value of correlation $r = 0.83$, for children $r = 0.80$ (Fig.~3).
Further, taking into account the observed correlations we
carried out  regression analyses of  the examined data.  The
linear regression equation were built to predict the state of
the RS  of the Kiev residents in terms of ``respiratory
diseases'' (RD) for adults and ``asthmatic bronchitis'' (AB) for
children depending on the annual peak concentrations of ozone
in their districts.

A prognostic study of the epidemiological situation with respect
to RD pathologies in Kiev in 2002--2006 was carried out with the
involvement of ground-level ozone measurements obtained at the
N.N. Grishko National Botanic Garden of  the National Academy of
Sciences of Ukraine, the satellite data (Aura-OMI), as well as
the modeled ozone concentrations for Europe (including Ukraine)
obtained by of the Rhenish  Institute for Environmental Research at the
University of Cologne.  We assumed that these data characterize
the ozone concentrations averaged over city in 2000--2007.  On
the base of these data we forecasted the overall incidences of
respiratory diseases (adults), and asthmatic bronchitis
(children).

\begin{table}[t]

{Table. Surface ozone impact on diseases of the respiratory system  
of population in Kiev  during 2000--2007. Comparison of the predicted
respiratory system  diseases   with the  medical
statistics (MS) data. Error of the  prognoses  is given in brackets} 

\begin{center}{

{ \small

\begin{tabular}{ccccccccc}
\hline
data source& 2000 & 2001 & 2002 & 2003 & 2004 & 2005 & 2006 & 2007 \\
\hline


Ozone concentration (ppb)    \\

Rhenish model & --    & --    & 75.0 & 75.0 & 55.0 & 78.5 & 71.5 & 90.5 \\
Ozonometer   & 56.9 & 74.4 & 75.5  & 71.1  & 67.9  & 84.2  & 77.0 & 91.5 \\
 Aura-Omi & --  & --  & --  & --& -- & 63.0 & 61.9 & 74.0 \\

\hline
       Respiratory diseases    \\

MS data & 2996 & 3093 & 3236 & 3534  & 3391 & 3561 & 3600 & 3836 \\
(per 100 000 adults)  &   &   &   &   & & & & \\

Prognoses with &  --   & -- & 3886  & 3884 & 2902 & 4058 & 3714 & 4649 \\
Rhenish model  &  & & (20\%) & (10\%) & (14\%)& (14\%) & (3\%) & (21\%) \\

Prognoses with & 2996  & 3857  & 3911  & 3694 & 3537 &4339 & 3985  & 4669 \\
Ozonometer  & & (25\%)  & (21\%)  & (4.5\%)  & (4.3\%)  & (22\%)  & (11\%)  & (22\%)  \\

Prognoses with   & --  & --  & -- &--  & -- & 3295 & 3241 & 3837 \\
 Aura-Omi  &--  &--   & -- & -- & -- & (7\%) & (10\%) &(0.003\%) \\

\hline
Asthmatic bronchitis  \\

MS data  & ~7.37 & ~8.39 & ~8.75  & ~9.2 & ~7.12  & ~9.2 & ~8.75 & --  \\
(per 1000 children) &   &   &   &   &  &  &  & \\

Prognoses with  & -- &--  & 9.7  & 9.7  & 9.3 & 10.1 & 9.0 & -- \\
Rhenish model   & & &  (11\%) &  (5.4\%) &  (24\%) &  (11\%) &  (3\%) & \\

Prognoses with & 7.69 & 9.9  & 11.  & 9.2  & 8.8 & 10.9 & 9.95& -- \\
Ozonometer & (4\%) & (18\%) & (26\%) & (0.0\%)& (6\%) & (18\%) & (14\%) & -- \\

Prognoses with & -- & -- & -- & -- &--  & 8.8 &  8.3 &--  \\
Aura-Omi & --& -- & -- & -- & -- & (8\%) &  (5\%) & -- \\
\hline
\end{tabular}
} }
\end{center}
\end{table}

Comparison of the results of prediction and health statistics
for 2000--2007 in Kiev (Table) showed that forecasts constructed
on the simulation results, in general, was confirmed. Thus, the
correlation coefficient for "Respiratory Diseases" (adults) was
0.76 for the ozone concentrations according to the simulation
and 0.72 for the data obtained by UV-ozone analyzer (TECO-49C).
According to satellite data Aura-OMI (2005--2007) for the same
indicators were obtained sufficiently close to the actual values
of assessing the incidences of pathologies of RS. Note also that
if for the central and southern Europe the peak ozone
concentrations occurred in 2003 followed by a slight decrease,
then for Eastern Europe this did not happen probably due to the increase
of road transport with old systems of fuel combustion, which are
prohibited in the EU countries.

It should be noted that the World Health Organization (WHO) 
\cite{WHO05} 
pointed out as especially dangerous for human health the four
air pollutants: particulate matter, ozone, nitrogen dioxide, and
sulfur dioxide. If ozone directly increases the risk of
RS diseases, the other three pollutants
increase the effect of ozone. As products of vehicle
and industrial enterprises emissions they also participate in the
ozone formation  as its precursors. In 
\cite{Lip_89} 
it was reported a
significant strengthening of the ozone effects on the functional
changes in the health of people in conjunction with other
adverse environmental factors. The recommendations of the WHO in
2005 
\cite{WHO05} 
lowered  maximum 8-hour average
concentrations of surface ozone from 60 to 50 ppb (120 to 100 mg
m$^{-3}$) due to the completion of knowledge of the
impact of ozone on health in epidemiological studies.

We calculated the root-mean square error $\sigma$ of one prediction
for each row in Table. For  ``respiratory diseases'' of
adults it was 15.0\% according to the  simulation of ozone
concentrations and 16.5\% according to the ozone measurements.
According to the rule 2$\sigma$ (i.e.  95\%
probability) we can assume all prognostic values as significant
ones. The correlation coefficient of medical statistics data and
forecasts according to ozonometer was 0.78. For diseases of
asthmatic bronchitis for children standard errors of forecast is
12.46\% on the base of the simulation and 14.98\% according to
ozonometer data.

Thus, we can assume all prognostic values as significant at the specified
forecast accuracy. Correlation coefficient of health statistics
for asthmatic bronchitis in children and forecast with
ozonometer data was 0.76. Note that the average ozone concentration
for the city of Kiev  based on the modelling (UAM-V) for
ozone episode in 2000 coincided with the value obtained by
UV-ozone analyzer at the National Botanical Garden. Therefore,
the use of the measured concentrations of ground-level ozone as
the average values for the city is justified for predicting RS
diseases.

Table shows that almost all the forecast data obtained on
the base of the Rhenish model and  the ground-based ozonometer data exceed
the health statistics data. There can be several reasons and among
them there are the reducing number of appeals to district health
centers (one part of the population prefers to be treated in
private clinics, and the other, needy, is self-treated). Also it
is appeared more effective medications, in particular, for
allergic asthmatic bronchitis.

\section{Conclusion}
The problem of forecasting of individual risk of the
ground-level ozone harmful impact on human health is
particularly acute in Ukraine, since the effective response from
government agencies to environmental problems tend to lag far
behind the needs of the population.

Our results suggest that the establishment of a monitoring
network of ground-level ozone measurements in the cities of
Ukraine, expanding the database with respect to ground-level
ozone pollution of the atmosphere of Ukraine cities, as well
as the statistical base for health and specific diseases of the
population of these cities is  a necessary condition for the
study of harmful effects of ozone on human health in Ukraine. As
a result of this work it became apparent that the modelling of
ground-level ozone in urban areas should be repeated at least
every 3--4 years to search for correlations between the
ozone concentrations  and RS morbidity of the population of
region taking into account rates of development of the cities,
as well as changes in the quantitative characteristics of ozone
dependent RS pathologies. Such studies combined with satellite
data (in particular, the data Aura-OMI) and modelling offer
opportunities for risk prediction of
harmful effects of ozone on the health of the population of
the cities in Ukraine.

{\it Acknowledgements.} This work
was partially supported by the Space Agency of Ukraine.

\end{document}